\documentclass[pdflatex,sn-mathphys-num]{sn-jnl}

\usepackage{graphicx}%
\usepackage{multirow}%
\usepackage{amsmath,amssymb,amsfonts}%
\usepackage{amsthm}%
\usepackage{mathrsfs}%
\usepackage[title]{appendix}%
\usepackage{xcolor}%
\usepackage{textcomp}%
\usepackage{manyfoot}%
\usepackage{booktabs}%
\usepackage{algorithm}%
\usepackage{algorithmicx}%
\usepackage{algpseudocode}%
\usepackage{listings}%
\usepackage{csquotes}
\usepackage{xurl}
\usepackage{tabularx}
\usepackage{booktabs}

\theoremstyle{thmstyleone}%
\theoremstyle{thmstyletwo}%

\theoremstyle{thmstylethree}%

\unnumbered

\begin{document}

\title[Retrospective on the Design and Implementation of the Milestone-Based Fusion Development Program of the
U.S. Department of Energy]{Retrospective on the Design and Implementation of the Milestone-Based Fusion Development
Program of the U.S. Department of Energy}


\author*[1,4]{\fnm{Scott C.} \sur{Hsu}}\email{scott@lowercarbon.com}
\author[2,4]{\fnm{Samuel E.} \sur{Wurzel}}\email{sam@fusionenergybase.com}
\author[3,4]{\fnm{Narayan S.} \sur{Subramanian}}\email{nss2130@columbia.edu}

\affil*[1]{\orgname{Lowercarbon Capital}, \city{Los Angeles}, \state{CA}, \postcode{90034}, \country{USA}}
\affil[2]{\orgname{Fusion Energy Base}, \city{New York},  \state{NY}, \postcode{10003}, \country{USA}}
\affil[3]{\orgname{Center on Global Energy Policy, Columbia University}, \city{New York}, \postcode{10027}, \state{NY}, \country{USA}}
\affil[4]{This is an account of work performed while the authors, listed with their present affiliations, were formerly with the U.S. Department of Energy. This paper represents the views of the authors and does not necessarily represent the views of the U.S. Government nor their present affiliations}


\abstract{This paper provides a detailed account of the origins, objectives, design, and implementation of the {\em Milestone-Based Fusion Development Program}, a public-private-partnership (PPP) program launched by the U.S. Department of Energy's
Office of Science in September~2022. The purpose
of this program is multi-faceted, with policy, scientific and
technological (S\&T), and commercialization objectives. Tactically, the program supports privately funded fusion companies in closing S\&T gaps toward delivering preliminary engineering designs of their demonstration fusion plants (aka ``fusion pilot plants''). The program is a key element of a shift in U.S. fusion strategy, initiated in 2022, to accelerate fusion commercialization by leveraging PPPs. The design of the Milestone Program was underpinned by its authorizing legislation and the U.S. National Academies report {\em Bringing Fusion to the U.S. Grid} (2021), and drew inspiration from other PPP programs discussed in this paper. Finally, the paper
provides the authors' perspectives and reflections on the strengths/weaknesses
of the program as implemented, implementation barriers, and lessons learned, with the
intent to inform future PPP programs in fusion and other energy/technology disciplines.}




\maketitle

\section{Origins}
\label{sec:origins}

The origins of the {\em Milestone-Based Fusion Development Program} \cite{FOA}, henceforth ``Milestone Program,'' of the U.S. Department of Energy (DOE)
can be traced to c.~2018 when the nascent Fusion Industry Association (FIA) started advocating for it with DOE and Congressional leaders. The FIA was inspired by the success of the earlier NASA Commercial Orbital Transportational Services (COTS) program \cite{COTS}, which
realized significant savings to the U.S. taxpayer via a milestone-based approach to developing new low-earth-orbit launch vehicles in partnership with the private sector. The NASA COTS program is credited with enabling the commercial space
launch industry, now dominated by U.S. companies like SpaceX (a former COTS participant).

Another aspect of the FIA's advocacy for the Milestone Program was the growth
in private investments into fusion companies in the late 2010s, with multiple fusion companies each starting to raise more than \$100 million, which
was sufficient to enable privately led research and development (R\&D) activities and experimental facility construction on par or exceeding that of publicly funded fusion institutions. Importantly, because fusion companies had increasing private capital from
a growing mix of investors, companies were becoming
more interested in alternative federal-funding mechanisms, such as Other Transaction (OT) agreements (much more on this later in the paper), that could enable
more industry-friendly terms and conditions relative to FAR (Federal Acquisition Regulation) contracts. In theory, this would attract wider industry
participation in public-private-partnership (PPP) programs, allow private
companies to move faster, and catalyze greater private investments, while providing
beneficial non-dilutive federal funding, validation, and partnerships. Despite the rapid growth of private investments into fusion companies in the late 2010s, it was not
as widespread as it is today, and the technical validation of a DOE Milestone Program was likely more important to
the leading fusion companies then compared to today.

During the same time period (late 2010s), there was also growing momentum and interest from the broader U.S. fusion research community to shift the U.S. fusion energy program from a purely scientific research program toward a timebound, mission-oriented program to realize a demonstration fusion plant. DOE Advanced Research
Projects Agency-Energy (ARPA-E) fusion programs played an influential role in stimulating these conversations and
bringing them into the mainstream \cite{nehl19,hsu25}. Several expert studies during this time period, which led to consensus reports released by the
U.S. National Academies of Sciences, Engineering, and Medicine (NASEM) \cite{NASEM19,NASEM21} and the DOE Fusion Energy Sciences Advisory Committee (FESAC) \cite{FESAC20}, reflected this evolving sentiment. Selected headline recommendations from
these reports include:
\begin{itemize}
\item The United States should start a national program of accompanying research and technology leading to the construction of a compact pilot plant that produces electricity from fusion at the lowest possible capital cost \cite{NASEM19}
\item (The DOE should) expand existing and establish new public-private partnership programs to leverage capabilities, reduce cost, and accelerate the 
commercialization of fusion power and plasma technologies \cite{FESAC20}
\item For the United States to be a leader in fusion and to make
an impact on the transition to a low-carbon emission electrical system by
2050, the Department of Energy and the private sector should produce net
electricity in a fusion pilot plant\footnote{As described in \cite{NASEM21}, a fusion pilot plant (FPP) should provide plant developers and owner/operators with the information needed to assess the economic attractiveness and
the role of fusion in the marketplace. An FPP should demonstrate a significant amount of net fusion electricity (e.g., $>50$~MWe) for $>3$ continuous hours (i.e., phase 1b of the NASEM report) with a timely path to one full power year (i.e., phase 2 of the NASEM report), at a total capital cost that can attract private funding.} in the 2035--2040 timeframe \cite{NASEM21}.
\end{itemize}

\begin{figure}[h]
\centering
\includegraphics[width=3.1truein]{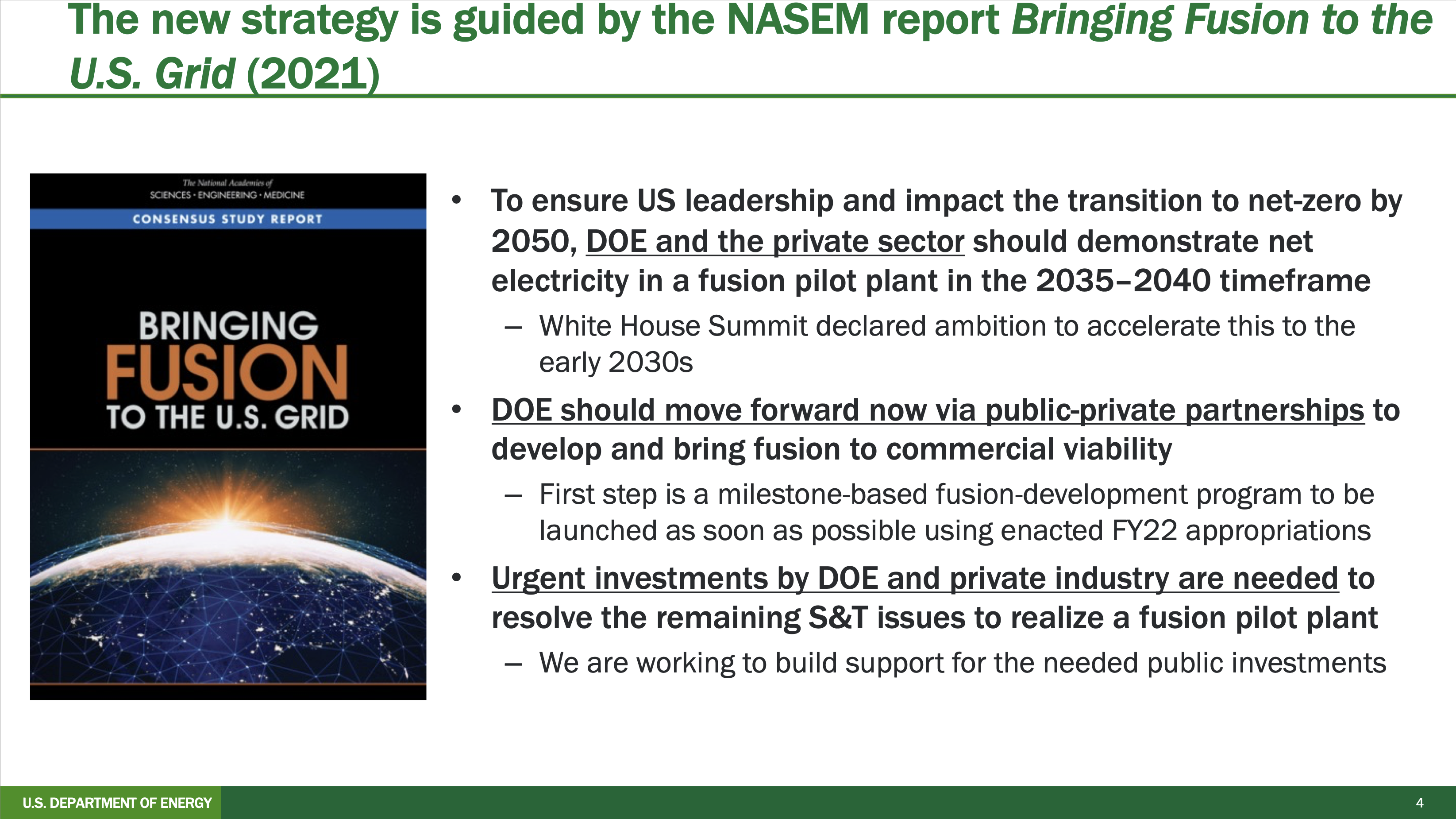}
\caption{Context for launching the Milestone Program from
the first author's presentation \cite{hsu22} at the DOE {\em Workshop on Fusion Energy
Development via Public-Private Partnerships} \cite{PPP-workshop} in June~2022 following
the White House Summit of March~2022.}\label{fig1}
\end{figure}

The convergence of advocacy by the FIA, support of commercialization-oriented fusion R\&D by ARPA-E, growing alignment from the broader U.S. fusion community as expressed
through consensus reports, increasing private investments into U.S.-based fusion companies, as well as support from senior DOE
leadership and several members of Congress, resulted in legislation that authorized the creation of the Milestone Program in the Energy Act of 2020 (later re-authorized in the CHIPS and Science Act of 2022). Notably, the Energy Act of 2020 augmented the mission
of the DOE Fusion Energy Sciences (FES) Program in the Office of Science (SC) with supporting ``the development of a competitive fusion power industry in the U.S.''
Further background on this history and the shift in U.S. fusion strategy toward accelerating fusion commercialization via
PPPs, initiated by the White House Office of Science and Technology Policy (OSTP) in late 2021, is given in \cite{hsu23}. The shift was formalized and publicly announced in a March~2022 White House Summit {\em Developing a Bold Decadal Vision for Commercial Fusion Energy} (BDV) \cite{BDV-fact-sheet,BDV-readout}, which provided
the necessary final catalyst (see Fig.~\ref{fig1}) for the DOE to move ahead with designing and implementing the Milestone Program.

Following are selected key elements of the authorizing legislation for the Milestone Program:
\begin{itemize}
    \item Program shall be established using DOE's OT authority [42 U.S.C. 7256(g)]
    \item Projects must meet particular technical milestones before a participant is awarded funds by the DOE
    \item Purpose of the program shall be to support the development of a U.S.-based fusion power industry through the research and development of technologies that will enable the construction of new full-scale fusion systems capable of demonstrating significant improvements in the performance of such systems, as defined by the Secretary, within 10 years of the enactment of this section
    \item Project proposals shall be evaluated based upon its scientific, technical, and business merits through a peer-review process, which shall include reviewers with appropriate expertise from the private sector, the investment community, and experts in the science and engineering of fusion and plasma physics
    \item Total authorized funding from fiscal years (FY) 2021 through 2027 (7 years) is \$415 million.
\end{itemize}

The Milestone Program received its first appropriated funds in FY 2022 and has been appropriated so far a total of \$200 million over 5 years through FY 2026 (see Table~\ref{table1}).

\begin{table}[h]
\caption{Enacted funding by year to the Milestone Program.}\label{table1}%
\begin{tabular}{@{}cc@{}}
\toprule
 Fiscal year & Enacted budget (\$M)\\
\midrule
2022 & 25 \\
2023 & 25 \\
2024 & 40 \\
2025 & $\ge40$\footnotemark[1] \\
2026 & 70 \\
\midrule
Total & 200\\
\botrule
\end{tabular}
\footnotetext[1]{No public information available, based on FY24 amount due to full-year continuing resolution in FY25.}
\end{table}

\section{Objectives and Design}
\label{sec:objectives}

The Milestone Program was formulated with multi-faceted objectives to fulfill the statutory requirements of the authorizing legislation (see immediately above), to
support the policy shift of the BDV, and to implement key NASEM-report recommendations \cite{NASEM21}.

The policy objective was to support the development of a U.S. fusion industry (consistent with the authorizing legislation)
and to accelerate fusion commercialization via PPPs that leverage and further catalyze private investments into U.S. fusion companies. A working thesis was that DOE's technical validation of mutually negotiated 
company milestones, as well as non-dilutive funding (at meaningful levels), would support private capital formation by the companies in the program
and thereby accelerate fusion commercialization.
The S\&T objective was to support fusion
companies in closing S\&T gaps (especially by encouraging and facilitating collaborations
among private companies, national laboratories, and universities) toward 
the realization of viable preliminary engineering designs of industry-led FPPs, consistent with or exceeding the requirements laid out in \cite{NASEM21}  (note that the BDV aimed for FPP operation by the early 2030s compared to the NASEM recommendation of the late 2030s).
The commercialization objective was to incentivize and support fusion companies in making 
progress in broader, largely non-S\&T areas, including but not limited to
capital formation, siting, regulatory, supply chains, byproduct-materials disposition, public engagement, workforce development, and market identification/preparation, which are all part of the DOE Adoption Readiness Level (ARL) framework~\cite{ARL}. The S\&T and commercialization objectives
were underpinned by Table~5.1 in the NASEM report \cite{NASEM21}.

The Milestone-Program Funding Opportunity Announcement (FOA)\footnote{Now known as Notice of Funding Opportunity (NOFO).} \cite{FOA}, which was released
on September 22, 2022, summarized the program's key objectives and eligibility
requirements as follows
(p.~2 of the FOA):
\begin{displayquote}
{\em This FOA invites applications for a new milestone-based fusion development program (as authorized in the Energy Act of 2020), which is a key component of the bold decadal vision to accelerate fusion energy RD\&D (research, development, and demonstration) in partnership with the private sector. Applications may be submitted for applied R\&D to resolve scientific and technological issues toward the successful design of a fusion pilot plant (FPP)\@.

It is expected that all applications will be led by the private sector with key partners that could include DOE national laboratories, academic institutions, non-profits, and other organizations/entities. Applicant teams should have a demonstrable range of technical and non-technical expertise needed for fusion energy R\&D and eventual demonstration and commercialization. Non-Federal financial commitments will be required for all applications.}
\end{displayquote}


Program design was informed by the above requirements and objectives, as well as (1)~community responses to a Request-for-Information (RFI), 
posted on April 20, 2020, on {\em Cost-Sharing Partnerships With the Private Sector in Fusion Energy} \cite{RFI}
(note in particular the response from FIA and its members~\cite{rfi-fia}), (2)~materials and discussions from a DOE
{\em Workshop on Fusion Energy Development via Public-Private Partnerships} held on June 1--3, 2022 \cite{PPP-workshop}, and (3)~extensive conversations with program officials and awardee/company executives of
the NASA COTS program, as well as program officials of the DOE Advanced Reactor Demonstration Program (ARDP)\@.

Based on a synthesis of all the aforementioned inputs/requirements, the following elements significantly guided program design:
\begin{itemize}
\item The project deliverable at the end of five years should be a successful FPP preliminary
design review (based on the NASEM report \cite{NASEM21}) or a significant performance improvement of the awardee's fusion concept (based on the authorizing legislation)
\item Eligibility was limited to for-profit domestic entities (or the subsidiaries/affiliates of for-profit foreign entities that are incorporated in the U.S.) as the primary award recipient
\item All awards were to be made under DOE's OT authority (per the
authorizing legislation) and therefore via 10 CFR 603--Technology Investment Agreements (TIAs)~\cite{TIA}, which at the
time was DOE's sole mechanism for making OT-based awards\footnote{DOE later recognized broader OT authority
under 42 U.S.C. 7256(a) as part of its January 2, 2025 rulemaking, as codified in 2 CFR 930--Other Transaction Agreements \cite{OT}.}
\begin{itemize}
\item Required awardees
to provide more than 50\% of the total project cost (TPC) via non-federal funding sources
\item Option to negotiate many terms and conditions that are standard in
       FAR contracts, especially those relating to intellectual property (IP), cost
        accounting standards (CAS), and reporting requirements
\item Option for awardees to receive federal fixed payments upon milestone completion
(thereby avoiding CAS), which
necessitated offering ``fixed-support'' TIAs as the award mechanism (10 CFR 603.300--315)
\end{itemize}
\item The program should support and evaluate S\&T, business, and
commercialization viability with equal weighting.
\end{itemize}

\section{Application Process}
\label{sec:implementation}

Immediately following
the June 2022 workshop \cite{PPP-workshop}, an integrated project team (IPT) (``Milestone FOA team'') was formed to develop, draft, and
release the FOA\@. The same IPT would later negotiate the terms of and make the awards. While the DOE FES program in SC
had programmatic and budget authority over the Milestone Program, the program and the FOA
were a priority and tracked at the highest levels of the DOE\@. Because of
the broad mandate for the program to support and evaluate S\&T, business, and commercialization milestones through the use of DOE's OT authority, the Milestone IPT included members from across the DOE (including the Office of the Under Secretary for Science and Innovation,
SC, ARPA-E, Golden Field Office, and the Office of Economic Impact and Diversity) to
bring expertise/perspectives spanning all these different areas.

The importance cannot be overstated of having
had senior DOE leadership support (particularly from the Undersecretary for
Science and Innovation and the leadership of the Office of Science) in prioritizing DOE resources and assigning experienced career staff for organizing/holding
the workshop and developing/releasing the FOA\@. What would have typically taken a year or more (organizing/holding the workshop and drafting/releasing a significant, first-of-a-kind FOA) took under five months. The FOA was announced on September 22, 2022
by then DOE Deputy Secretary David Turk at a fusion showcase event held as part of the Global 
Clean Energy Action Forum in Pittsburgh, PA \cite{FIA-announcement}.
Later, 
DOE recognized Milestone IPT members, as well as a few others who accelerated final approvals
in time for the announcement, with a ``Special Act Award for the Milestone Fusion Program FOA Announcement Team.''

\subsection{Technical Requirements of the FOA}
\label{sec:tech}

Key requirements for applications to the Milestone Program were described in the FOA \cite{FOA} as follows (pp.~3--4 of the FOA):
\begin{displayquote}
{\em
    Applications shall propose a series of milestones toward realizing an FPP\@. The initial project deliverables, by 18 months after award, are FPP preconceptual designs and technology roadmaps, with the understanding that funding for subsequent milestones up to a total period of performance of five years will be contingent upon meeting early milestones and the availability of appropriated funding...

    As discussed in the NASEM report \cite{NASEM21}, a conceptual design of an FPP leading to preliminary and final engineering designs will be required. Table 5.1 in the NASEM report lists technical and non-technical actions that should be completed to enable conceptual, preliminary, and final designs. A preconceptual design would need to address the same issues as the conceptual design but at lower levels of fidelity and with greater uncertainties. As also discussed in the NASEM report, a technology roadmap identifies in detail the required critical-path R\&D, including any intermediate test facilities, and focuses on the advances required for a particular FPP conceptual design.
    
Each application shall select one of two Tiers as described below...:
\begin{itemize}
\item Tier 1: Applications shall articulate a plausible path and proposed milestones leading to a successful preliminary design review (PDR) by the late 2020s, of an FPP that can begin operations by the early 2030s. Significant commitment of non-Federal resources is expected for all Tier 1 applications...
\item Tier 2: Applications shall articulate a plausible path and proposed milestones leading to a significant improvement, as defined quantitatively by the applicant, in the fusion performance (including the physics basis and required enabling materials/technologies) of their proposed FPP concept by the mid/late 2020s. Applications shall further describe how the improved performance may lead to a successful PDR, by the early 2030s, of an FPP that can begin operations by the late 2030s. Significant commitment of non-Federal resources is expected...
\end{itemize}

Milestones should reflect critical-path
scientific/technical, business/financial, commercialization, and EJ/DEIA-related\footnote{Energy justice and diversity, equity, inclusion, accessibility.} (including
socio-technical\footnote{A ``socio-technical'' milestone is one where consideration of EJ and/or public acceptance wholly or partially motivates demonstration of a particular technical requirement, e.g., materials specifications to avoid or minimize
radioactive byproduct materials, a blanket/tritium-system design to minimize tritium site inventory, etc.}) deliverables. Provide estimated total and requested federal funding amounts
for each milestone. Applicants are strongly encouraged to include particular milestones to
address the following scientific and technical requirements, if applicable within the 5-year period
of performance:
\begin{itemize}
\item Achieve plasma conditions (i.e., the required Lawson parameter $n\tau$ and ion temperature
$T_i$) \cite{wurzel22} needed for a significant improvement in equivalent (i.e., using D-D instead of the
actual commercial fuel) scientific energy gain up to and beyond $Q>1$
\item Achieve scientific energy gain $Q>1$ using the chosen commercial fuel cycle
\item Heat-exhaust and plasma-facing-component (PFC) solutions for the FPP
\item Sustainable fuel-cycle solution, including blanket and tritium processing if applicable, for
the FPP
\item Actuators and key enabling technologies.
\end{itemize}
    }
\end{displayquote}

Applicants were advised that S\&T assertions should be supported by experimental data and theory/modeling-based analysis to the
extent possible, with citations to peer-reviewed publications if available, and were asked to address in their applications a more detailed list of  considerations in formulating their proposed milestones (see pp.~5--8 of the FOA \cite{FOA}).

The Milestone Program FOA \cite{FOA} covered a period-of-performance of five years, with 
project deliverables as stated above. However, the intent was always for the program to continue beyond five years as long as there were credible companies making meaningful progress (and continuing to raise private funding) toward FPP preliminary engineering designs. The program was designed so that a ``demonstration tier'' could easily be added in the future to support FPP final engineering
design and construction for companies that pass their FPP preliminary engineering design reviews.

\subsection{Administrative Requirements of the FOA}
\label{sec:admimistrative}

Administrative requirements and supporting information filled up the rest of the FOA (pp.~10--80), including information
on awards, eligibility, application/submission, application reviews, award administration, and supplementary materials. Following
are some aspects of the FOA that deviate substantially from standard DOE FOAs that award FAR-based contracts:
\begin{itemize}
\item Awards would be made under DOE's OT authority via TIAs, noting that this would allow greater flexibility in tailoring
terms and conditions related to audits, CAS, and IP rights
\item Applications were accepted only from lead entities that are for-profit domestic entities or subsidiaries/affiliates of for-profit foreign entities that are incorporated in the U.S.
\item If a DOE/NNSA National Laboratory or that of another federal agency is part of a teaming arrangement, the Laboratory is expected to perform work under a
contractual agreement with the lead entity (as permitted under the Laboratory's Management and Operating contract with the DOE) rather than via a direct work authorization from the DOE
\item DOE would have substantial involvement in the award by being engaged in an advisory capacity for work performed under the award
\item Anticipating that most or all awards would be made under ``fixed-support'' TIAs (i.e., fixed payments of a pre-negotiated
amount upon verified milestone completion), cost sharing (as required by Sec.~988 of the Energy Policy Act of 2005) was waived to eliminate or reduce CAS requirements, although awardees were still expected to provide non-federal resources exceeding half the TPC
\item Detailed line-item budget breakouts were not required for applicants requesting fixed-support TIAs as the award mechanism; instead applicants only needed to provide a concise justification and basis for the total cost to complete each milestone
\item In addition to S\&T viability (30\%), the criteria for merit review included commercialization (30\%) and business/financial (30\%) viability, as well as EJ/DEIA (10\%) (see pp.~29--30 of the FOA \cite{FOA}).
\item Certain categories of data produced by awardees in this program would be protected from public disclosure for 20 years, and
awardees could request extended protection for up to 30 years
\item DOE anticipated issuing a class patent waiver for any domestic large business that provide at least 20\% cost share, where
the domestic large business would be able to elect title to their subject inventions similar to the right provided to domestic small
businesses, educational institutions, and nonprofits by law.
\end{itemize}

\subsection{Merit Review and Selections}
\label{sec:review}

As directed by the Milestone Program's authorization language and to support its policy objectives, the merit review of 
applications (following a pre-application phase that screened only for 
programmatic responsiveness, see p.~15--17 of the FOA~\cite{FOA}) included reviewers (including both federal and non-federal) with appropriate expertise from the private sector, the investment community, and 
experts in the science and engineering of fusion and plasma physics. Reviewers from DOE National Laboratories and universities
were required to sign a statement of confidentiality, and reviewers from the private sector were required to sign a non-disclosure
agreement.

Each application was assigned up to 10 reviewers spanning the various required areas of expertise, taking care to avoid 
conflicts of interest (which was challenging given the finite size and high connectivity of the U.S. fusion research community).
In support of U.S. competitiveness objectives, the Milestone Program
uses only U.S. persons as expert reviewers except in very limited cases
when absolutely necessary.

Reviewers first read the full applications assigned to them and provided written comments on the applications' strengths
and weaknesses relative to the
merit-review criteria (pp.~29--30 of the FOA \cite{FOA}). Based on the written comments as well as the program policy factors
(see p.~31 of the FOA \cite{FOA}), 
DOE program staff selected finalists, who were provided a minimum of 2 weeks to prepare
for 2-hour oral interviews conducted by videoconference.
In the first hour, presenters gave an overview of the proposed project and addressed DOE questions (based in part
on reviewer comments) that were sent in advance. The second
hour was a general question-and-answer period. The interviews were attended by up to 5 members of the presenting team
(2 presenters plus 3 additional team members to answer questions as needed), DOE program staff, and non-federal reviewers.

Based on all the information obtained throughout the merit-review process, DOE program staff identified eight highly
meritorious applications that we felt were essential to fund with the available budget at the time (\$50 million for
the first 18-month period-of-performance), with two additional applications deserving of funding if more funding was available.
Although the FOA indicated that DOE anticipated making between three and five awards (see p.~11 of the FOA \cite{FOA}), two
of us (S.H. and S.W.)
advocated strongly to make awards to all eight of the highly meritorius applications for two key reasons: (1)~higher probability of
program success of delivering at least one viable preliminary design of an FPP by the end of the program, due to the considerable technical and non-technical risks faced by any single team and fusion approach, and (2)~to catalyze the greatest amount
of private-sector investment in fusion development. Because at least an order-of-magnitude higher funding
was needed for the program to fully meet its objectives (relative to the \$50 million that had been appropriate at the time of the
FOA release), we felt that initial dilution of funding to more companies was worth accepting to provide an opportunity for the most meritorious companies to organically emerge when higher levels of funding might be appropriated to the program. The question of ``downselection'' is discussed later in the paper.
Foreign-influence and research-security reviews of selectees occurred only after merit reviews were completed
instead of in parallel, and delayed announcement of selections by over a month. The eight companies (see Table~\ref{table:selections}) selected for award negotiations were
announced on May 31, 2023 via a livestreamed DOE virtual event \cite{selection-announcement}.

\begin{table}[h]
\caption{Inaugural selectees/awardees of the DOE Milestone Program. Cumulative obligated and paid amounts are from \texttt{https://www.usaspending.gov} 
as of September 10, 2026 with some further updates based on information from the awardees. Equity-investment amounts are as of September 10, 2026 from \texttt{https://www.fusionenergybase.com}, which is based on publicly available information and may not include undisclosed amounts. Implications of the small ratio of obligated amounts relative to total equity investments are discussed in the main text.}\label{table:selections}
\begin{tabular*}{\textwidth}{llcccc}
\toprule%
Selectee/awardee & Approach & Obligated & Paid & Approx.~Equity & Obligated/ \\
   & & (\$M) & (\$M) & Investments (\$M)& Investments (\%) \\
\midrule
CFS & Tokamak & 33.1 & 9.7 & 3,908 & 0.5 \\
  Xcimer Energy & Laser IFE & 13.5 & 8.5 & 165 & 8.2  \\
  Type One Energy & Stellarator & 9.5 & 3.0 & 175 & 5.4 \\
  Zap Energy & Z Pinch & 17.8 & 4.0 & 324 & 5.5 \\
  Realta Fusion & Mirror & 12.5 & 3.14 & 45 & 18.0 \\
  Thea Energy & Stellarator & 13.7 & 3.0 & 130 & 10.5\\
  Tokamak Energy & Tokamak & 7.5 & 3.0 & 274 & 2.7\\
  Focused Energy & Laser IFE & 7.5 & 2.15 & 267 & 2.9 \\
  \midrule
   Total & & 115.1 & 36.5 & 5,288 & 2.2  \\
   Total minus CFS & & 82.0 & 26.8 & 1,380 & 5.9 \\
\botrule
\end{tabular*}
\end{table}

\section{Award Negotiations}
\label{sec:negotiations}

Award negotiations brought significant challenges with respect
to exercising and implementing many of the flexibilities, as authorized in 10 CFR 603 \cite{TIA}, that supported the policy objectives 
of the Milestone Program. 
It must be emphasized that DOE had the authority to enter into TIAs since the Energy Policy Act of 2005. Through a Secretarial Working Group launched in 2022, DOE had just recognized broader statutory authority in its 1977 Organization Act to exercise OT authority. DOE had also just published a Guide to Other Transactions in 2023 \cite{DOEOTGuide23} along with a new training for AOs to obtain the necessary warrant to execute an OT agreement, noting that fewer than three AOs across the DOE complex had such a warrant at the time. Additionally, DOE was in the process of updating its TIA and general OT regulations, which was not complete until January 2025. All of this meant that while DOE leadership was ready to lean into the flexibilities in 10 CFR 603, its institutional experience negotiating and executing such agreements was severely limited, especially in a situation where private funding dwarfed the federal funding.
At the same time, 
many selectees/companies had unrealistic expectations that the flexibilities would translate to DOE waiving all IP rights or oversight in certain areas. This misalignment/chasm between the two sides at the start of award negotiations contributed to the significant delays to come.
Award negotiations consisted of two interconnected components: negotiation of milestones and negotiation of TIA terms and 
conditions. The latter presented the most significant challenges.

\subsection{Milestone Negotiations}
\label{sec:milestone-negotiations}

Milestones underpin the program. Well-chosen milestones satisfy competing objectives: they represent major de-risking and valuation inflection points
on a company's path to an operating FPP and beyond (including S\&T, business, and commercialization milestones), which favors fewer and more significant milestones (also reducing administrative burdens for both the DOE and companies), versus providing sufficient granularity for more frequent non-dilutive federal payments to a company along their development path, which favors a higher number of less significant milestones. 

There were a few milestones that were ``required'' of all selectees, such as proving capital runway for executing the milestones being 
negotiated,
FPP preconceptual designs, technology roadmaps, 
community engagements. Though the initial milestone negotiations focused on the first 18 months
of the program, a final milestone was required at the end of five years for either an FPP preliminary engineering design or a significant performance improvement of the selectee's
FPP concept (as negotiated). Selectees were asked to propose a set of milestones based on the required milestones and the milestones in their applications. The intent here was to allow selectees to align with their investor milestones as a starting point for milestone negotiations. Companies had the option of protecting the details of their 
milestones from public disclosure for up to 30 years, and were encouraged to publicly describe their milestones as well as milestone completions at a high level to
build momentum and maintain transparency. Peer-reviewed publications of S\&T advances were encouraged, in a manner that protects a company's IP, in order to build and maintain credibility.

Aside from the required milestones, the final negotiated milestones were largely customized for each selectee. They generally focused on near-term
S\&T priorities (e.g., plasma performance, enabling technologies such as magnets, lasers, or pulsed power, and blanket/fuel cycle technologies), business 
needs (e.g., private fundraising, key hiring, expanding operations, etc.), and commercialization (e.g., licensing/permitting, siting, community
engagements, workforce development, customer engagements/partnerships, etc.). Table~5.1 of \cite{NASEM21} was suggested as a guide
to inform milestone scope.

Milestone negotiations began a few weeks before public announcement of the eight selectees (May 31, 2023). The first pass of milestone negotiations proceeded relatively quickly
and smoothly over 1--2 months (aiming to match ARPA-E fusion programs \cite{hsu25}, for which the internal goal was to complete technical milestone negotiations within one month of announcement of project selections). Selectees were asked to propose an initial set of milestones, including estimates of total cost and federal-payment amounts per milestone, based on the 
content of their original applications and their awarded budget, and to propose milestones and completion criteria that significantly de-risked their S\&T, business growth, and commercialization pathways to an FPP and beyond. 

In general, the selectees proposed a very good set of initial milestones. Business and commercialization milestones
were typically straightforward, e.g., show proof of capital runway via bank deposits/statements, hold a certain number of community two-way engagements to inform/advance siting activities, submission of a licensing application, etc. S\&T milestone negotiations focused on
adding specificity to milestone-completion criteria. As done previously at ARPA-E, selectees were pushed to articulate required performance targets and measurements that demonstrate meaningful advances in technological readiness level (TRL) on the path to an eventual FPP (e.g., not just ``build and test a magnet or laser'' but identifying meaningful metrics and required performance targets for those metrics). Selectees were repeatedly asked the question: ``What
performance targets do you need to meet in order to feel that a significant S\&T risk has been retired and to feel confident to proceed in your FPP development path with acceptable S\&T risk?''

The Milestone Program aims to provide insight and not oversight (thanks to former NASA COTS program manager Alan
Lindenmoyer for this wisdom). There is a tendency for technical experts and federal program managers to want to overly dictate to a company the things that need to be de-risked, in what order, and to what extent. 
While this is understandable when taxpayer dollars are funding a large majority or all of the work, in the Milestone Program, companies are taking the upfront and predominant financial
risk (i.e., see last column of Table~\ref{table:selections}), and therefore it should be largely up to the companies (and
their investors) to decide what risk they are willing to take with 
respect to milestones and milestone completions. DOE (and its expert reviewers) should provide as much advice and insights as possible to the 
companies so they have the greatest chance of success under constrained budgets, and to give their boards/investors as much confidence as possible in the company's development path and progress.

Milestone negotiations were largely complete within 1--2 months following the announcement of selectees. However, due to where the negotiations eventually landed on TIA
terms and conditions many months later in the negotiations process (discussed next), another pass at the milestone negotiations was
needed for most of the selectees. Many companies reduced the scope of their initially negotiated S\&T milestones (as part of the Milestone Program) to avoid subjecting themselves
to certain IP- and reporting-related terms and conditions in the final TIA\@. Reducing the scope of S\&T milestones was counter to one of the policy objectives
of the program, which was to accelerate S\&T progress via broader public-private collaborations.
Opportunities remain to improve the TIA terms and conditions such that awardees embrace the broadest set of milestones and 
partnerships with publicly funded researchers and institutions.

\subsection{Negotiation of TIA Terms and Conditions}
\label{sec:TIA-negotiations}

Getting consensus on the initial terms and conditions of a draft TIA template, with industry input, prior to releasing the FOA would have been a good idea. This would have surfaced award-negotiations challenges earlier, but it may not have
ultimately saved any time because the associated growing pains were likely unavoidable. Regardless, it
was decided to release the FOA as soon as possible and develop the TIA template afterward to keep up the momentum following the White House summit in April 2022.

A key stumbling block was that, despite having general agreement among the Milestone IPT members on the policy objectives
of the Milestone Program (described in Sec.~\ref{sec:objectives}), the AOs and patent counsel on the Milestone IPT
commenced negotiations largely using the standard terms and conditions of FAR-based contracts under the assumption
that the TIA would receive swifter approval from the DOE Office of Management. Program staff (especially two of us, S.H. and S.W.) did not agree with this approach,
as it was foreseeable that selectees/companies and their board members would not accept these terms pertaining to a number of areas, discussed in turn in the subsections below. Nor were we empowered to override the AOs' authority.

As a result, the IPT expended six months of back-and-forth with the selectees
before conceding a stalemate, at which time the Office of the Secretary of Energy (one of us, N.S., became engaged
in the award-negotiations process at this time) was
approached by one of us (S.H.) to help move past the stalemate.
The AOs were compelled to be responsive to other policy directives (e.g., the Secretary's directives on promoting domestic manufacturing and research/supply-chain security) under their own DOE
chain of command, and ultimately this required intervention from the Office of the Secretary to provide guidance on which imperatives should prevail. In the following subsections, we discuss in detail how we worked past the stalemate.

An additional contributing factor to the stalemate was the relatively small federal contribution for the first 18 months of the program (see last column of Table~\ref{table:selections}). If the Milestone Program was funded adequately to allow for closer to 50\% federal share of total company spends, as allowed by statute for OT agreements, 
terms closer to those of FAR-based contracts might have been more palatable to at least some of the selectees as the starting point (not necessarily the
end point) for award negotiations (e.g., most of the DOE ARDP awards, which had much higher federal funding, used FAR-based contracts).
Even in the case of much
higher federal share, deviations from FAR terms would still be needed to support the policy objectives of the Milestone Program. Some of the selectees emphasized that the negotiations should not be based on FAR principles
regardless of federal funding amount, given that the Milestone Program was and is not intended to be an acquisition program.
Instead, it was and is intended to stimulate the growth of a U.S. fusion industry, which would be the 
primary value to the taxpayer 
rather than any specific IP or government capability generated as part of the program. This tension remains unresolved and warrants further consideration/discussion,
especially as the Milestone Program evolves toward potentially supporting demonstration-phase projects.

Due to the extreme challenges of the negotiations, and a strong desire by DOE program staff (not
necessarily the entire IPT) for all eight selectees to
receive the same final TIA terms and conditions, the eight selectees (which came to be known as the ``M8'') started to
converse with each other to ensure balanced negotiations and sharpen their primary asks.
While this was unprecedented for selectees of a DOE program, it was somewhat understandable considering the small ecosystem of fusion companies and, for some companies, their limited resources to hire independent government contracting and patent counsel. In effect, this meant the fusion companies negotiated as a bloc.
The AOs, patent counsel, and
their chains of command were very averse to coordination among the M8 during award negotiations. A few takeaways here are
that the coordinated M8 negotiating tactic was probably needed to ensure the best possible outcome under the circumstances,
but that the M8 coordination at the award-negotiations phase may not have been necessary if broad industry input was collected and meaningfully addressed in drafting a TIA template ahead of time.

\subsubsection{Reporting Requirements}
\label{sec:reporting}

Minimizing reporting burdens throughout a program's lifecycle is important to private companies, especially for early-stage, venture-backed companies with limited capital and small teams focusing on business growth and key S\&T de-risking activities. We aimed to limit
upfront and ongoing reporting requirements as much as possible to only the
information needed to evaluate whether a milestone was successfully completed, as permitted under 10 CFR~603.880(a). However, despite our best efforts to minimize or eliminate most of the reporting requirements, as permitted under 10 CFR~603.880--603.900, the reporting
checklist (included as part of the final award document package) ultimately was based on a standard DOE reporting checklist (DOE~F~4600.2) of nearly 20 pages in length with the following high-level requirements: (1)~management/progress reporting, (2)~S\&T reporting, (3)~closeout reporting, (4)~subject-invention and patent reporting, (5)~collaborating organizations, (6)~foreign connections, and (7)~current and pending support. 

Fortunately, the IPT was able to eliminate the most onerous requirement of detailed financial reporting based on CAS, which requires very granular budget breakouts by expenditure category. Normally, per 10 CFR~603.880(b), CAS-based financial reporting is required in order to verify specific non-federal cost-sharing amounts (e.g., at least 50\% for OT agreements per the Energy Policy Act of 2005). Therefore, the IPT executed a cost-share waiver (requiring the approval of a Senate-confirmed appointee) prior to the release of the
FOA to eliminate CAS while retaining a requirement in the FOA of at least 50\% non-federal contributions to the TPC (see p.~13 of the FOA~\cite{FOA})\@. This arrangement would not have been possible otherwise under 10 CFR 603.
As a result, for fixed-support TIAs, only high-level estimates and justifications for the cost of meeting each milestone were required (instead of line-item granularity) in order to confirm that federal payments were less than 50\% of the cost to complete a milestone.

However, many of the other reporting requirements presented a challenge during award negotiations, requiring finding the proper balance between minimizing reporting burdens for the 
selectees and collecting enough information to meet various statutory requirements. Of the requirements enumerated two paragraphs previously, (4)--(7) presented the greatest
challenges. There were two separate issues that exacerbated these reporting requirements. Firstly, there was the question of whether any reporting requirement should ``flow down'' to the primary recipients' collaborators and subcontractors, which were not receiving any federal funds directly. Ultimately the answer was ``yes'' but limited to the named institutions and key collaborators in the original application rather than all collaborators. Secondly, there was the question of the level of detail needed in what was reported to meet a requirement. In general, the TIA regulations (10 CFR 603) provided some degree of flexibility in interpreting these requirements, and ultimately it was the decision of the AO with input from IPT members. In general, the AOs preferred to adhere to standard practices of DOE
contracting as a starting point of negotiations rather than exercise the flexibilities afforded in 10 CFR 603. However, to the credit of the AOs and patent counsel, several flexibilities relating to IP and reporting terms were offered even prior to intervention by the Office of the Secretary. Further commentary is provided here:
\begin{itemize}
\item S\&T reporting: Peer-reviewed publication of S\&T results was encouraged whenever possible, especially as it relates to required advances in plasma 
performance, e.g., fusion triple product \cite{wurzel22}, toward $Q>1$ and beyond.
However, submission to a peer-reviewed publication was not a requirement
before conducting a milestone-completion review or render a decision. The only requirement for S\&T milestones 
is a written technical report with sufficient detail
for DOE and external reviewers to evaluate whether a milestone was completed. This
was to avoid delays due to the time it takes to prepare articles of sufficient
polish for submission to a peer-reviewed journal. The IPT and selectees were generally in agreement here.

\item Subject-invention and patent reporting: This was a major source of tension that also played a role in many selectees revising their milestones after the initial
milestones were negotiated. Invention reporting is a standard practice under normal financial-assistance contracts such as grants and cooperative agreements, where federal
funding typically supports 80--100\% of the TPC\@. However, given that selectees in the Milestone Program are providing a large majority of funding (see Table~\ref{table:selections}), they were not willing to accept standard invention reporting requirements. From the federal perspective, in the event that
the DOE needs to exert its negotiated
government-use license (more on this below), the DOE must have awareness of subject inventions (with sufficient detail) and patents developed with partial federal support under this program. Where the negotiations landed was to require subject-invention and patent reporting on particular technical milestones and within a predefined
technical scope for those milestones (as negotiated). The predefined scope size would be subjectively commensurate with the fractional share of the federal payment for that milestone.
Though this was a somewhat clunky solution, it enabled both sides to sign the TIA and is also a scalable solution if federal contributions go up in the future.

\item Collaborations, current/pending support, and foreign connections: These are typical reporting requirements for any federal application or award to identify potential conflicts of interest, duplication of federally funded work, and research-security and
foreign-influence risks. The main concern here was related to the burdens of ``flow-down''
reporting requirements to collaborator and subcontractors, with a compromise of limiting the reporting requirement to the original named institutions
and key collaborators in the project application.

\end{itemize}

\subsubsection{IP Rights}
\label{sec:IP}

Negotiation of IP rights was a significant challenge, given that negotiations started largely and inappropriately from the standard
IP provisions of DOE Federal Financial Assistance \cite{CDSB-115}. The standard IP provisions were inappropriate in this instance given that the selectees
were in some cases spending an order-of-magnitude more in private funding compared to their federal funding in the Milestone Program. 
Furthermore, as mentioned above, many selectees/awardees argued that building a U.S. fusion industry rather than the
generation of government IP rights or capabilities is the primary benefit delivered by this program, and therefore the government should attach the absolute minimum IP/data rights to awardees, regardless of federal 
funding level. This tension remains as an open question for present and
future stewards/participants of the Milestone Program (and others like it) to address.

Flexibilities on IP terms were explicitly allowed under 10 CFR~603.840--603.870 \cite{TIA}. For example, 10 CFR~603.840 states:
\begin{displayquote}
   ``(a) The contracting officer must confer with program officials and assigned intellectual property counsel to develop an overall strategy for intellectual property that takes into account inventions and data that may result from the project and future needs the Government may have for rights in them. The strategy should take into account program mission requirements and any special circumstances that would support modification of standard patent and data terms, and should include considerations such as the extent of the recipient's contribution to the development of the technology; expected Government or commercial use of the technology; the need to provide equitable treatment among consortium or team members; and the need for the DOE to engage non-traditional Government contractors with unique capabilities.

(b) Because a TIA entails substantial cost sharing by recipients, the contracting officer must use discretion in negotiating Government rights to data and patentable inventions resulting from the RD\&D under the agreements. The considerations in \S\S~603.845 through 603.875 are intended to serve as guidelines, within which there is considerable latitude to negotiate provisions appropriate to a wide variety of circumstances that may arise.''
\end{displayquote}

And 10 CFR~603.860(b) states:
\begin{displayquote}
``The contracting officer may negotiate Government rights that vary from the statutorily-required patent rights requirements described in paragraph (a)(2) of this section when necessary to accomplish program objectives and foster the Government's interests.''
\end{displayquote} 

The disagreements during early negotiations centered around data rights,
government's data/copyright and patent licenses, march-in-rights, and a number of other procedural clauses. 
While the selectees wished to eliminate most or all of these provisions (i.e., they essentially wanted to receive a federal
``prize payment'' upon the completion of each milestone), this was not
a viable option from the perspective of DOE patent counsel. This disagreement was a 
significant contributor to a multi-month delay/stalemate in the award negotiations, and we were only able to move forward once we engaged
the Office of the Secretary of Energy to receive policy guidance and directives on breaking the stalemate.

While we cannot reveal the precise IP provisions of the final negotiated and signed TIAs, as these are considered to be proprietary 
commercial information of the awardees, we summarize some key modifications relative to the standard DOE IP provisions:
\begin{itemize}
    \item Government's data rights are narrowed, including recipient's right to withhold computer software
    \item Government's data/copyright license was narrowed and explicitly excludes commercial use
    \item Government's right to subject inventions made under the award was narrowed, and DOE agrees not to
    exercise this right for a significantly extended period past the relevant milestone completion
    \item March-in-rights are replaced with a ``conditional license'' tied specifically to commercial deployment of the recipient's FPP, with longer cure periods and a compliance safe harbor for a significantly extended period past the relevant milestone.
\end{itemize}

Selectees still had concerns about the narrowed government's right to data and subject
inventions, given that many of them are interested in potentially becoming commercial suppliers of such technology to the government
for non-commercial use. Two of us (S.H. and S.W.) advocated strongly for inclusion of the safe-harbor clause (inspired by the NASA COTS award documents),
and in our opinion, this was a significant factor in allowing all parties to come to agreement on accepting the TIA\@. Nevertheless, in many instances,
the final TIA IP provisions still caused many of the selectees to reduce the scope of and/or eliminated certain milestones to avoid
entanglements with the government's IP rights for the potential subject inventions associated with those milestones. A compounding factor was that the eight selectees were at very different stages of fundraising and company building, and had different levels of resources to expend
on their own patent/IP advice, which also probably contributed to delays at various stages of the negotiation process. In the end, de-scoping of the milestones was counter
to the policy objective of the Milestone Program, which intended for the DOE to be a partner with fusion companies in broadly supporting their overall S\&T de-risking and development path toward viable FPP designs. To best support the headline policy objectives of the program
(i.e., building a competitive U.S. fusion industry and accelerating fusion commercialization by partnering with the private sector),
it is worth reconsidering the appropriate balance of IP rights that is commensurate with the ratio of private-to-federal funding.

\subsubsection{Foreign-Work Waivers and Research-Security Reviews}

There are standard procedures for requesting waivers for conducting R\&D work on foreign soil by DOE awardees and periodic reviews of potential
foreign influence on the awardees. There was (and still is) significant
pressure on federal agencies to address/mitigate research-security concerns through tight controls on foreign work by and influence on recipients of U.S. federal
funding. Most of these controls could not easily be waived by the DOE without drawing heavy scrutiny. The 
main issue for the award negotiations was the uncertainty in the timeline for arriving at a decision to approve a waiver request
and the administrative burdens of recurring research-security and foreign-influence reviews. Companies in the Milestone Program
are venture-backed companies trying to move as quickly as possible, and this translates to the need for predictable and rapid timelines in
securing waiver approvals (and pretty much all administrative decisions). The compromise was the addition of a clause in the TIA stating that DOE would review foreign-work waivers and issue a written
determination within 45 days of receipt of the waiver request. For initial and recurring foreign-influence reviews, a compromise was to limit
the review to the awardee and key collaborators in the original project application (as mentioned above).

\subsubsection{U.S.-Manufacturing Requirements}

U.S.-manufacturing requirements in the TIA are based on DOE's June 7, 2021 Determination of Exceptional Circumstances under the Bayh-Dole Act ``to further promote domestic manufacture of DOE science and energy technologies,'' as well as 10 CFR~603.875(c). The requirement is that any products embodying or made using a subject invention made under the award to be manufactured substantially in the U.S\@. This was a marquee policy (otherwise known as ``Invent it Here, Make it Here''), which had been championed by Secretary Granholm and later adopted by President Biden as Executive Order 14104 on July 28, 2023 and continues to be in effect. Hence, this naturally became a difficult policy from which to request a waiver. 
Nevertheless, the awardee (or any bound downstream entity) may request a waiver/modification of the above requirement if it can show, with quantifiable data, that domestic manufacturing is not commercially feasible and that a waiver serves U.S./public interests. As with above, a key concern during
award negotiations was the timeline for determining whether a waiver request would be approved. Again, the compromise was the addition
of a clause that DOE would act within 45 days of the submission of the waiver request (with a 10-business-day acknowledgment where practicable). If DOE does not object within 45 days, the waiver would be deemed approved.

A higher-level question that we did not attempt to address during award negotiations is the appropriateness of imposing U.S.-manufacturing requirements at the R\&D and innovation phase
on an industry that is very much aiming in the future to construct/site fusion plants globally around the world,
which could also require foreign-based manufacturing of fusion-plant components for maximum economic competitiveness.
How these policies impact the ability for frontier technologies to scale quickly warrants broader consideration by policymakers.

\section{Initial Program Execution}

On June 6, 2024, almost exactly a year after the Milestone selectees were announced on May 31, 2023, a second White House event \cite{BDV2024} 
was held to mark two years of the {\em Bold Decadal Vision for Commercial Fusion Energy}. As part of the event, it was announced that
all eight Milestone awardees had signed the TIA\@. Indeed, the event provided a hard deadline for completing the award negotiations.

By the time the TIAs were signed, many companies had actually completed or were close to completing many of their early milestones, and thus we sprinted straight into program execution in terms of receiving milestone-completion reports and organizing 
the first milestone reviews for several companies. In some cases, milestone completion was simple verification of documentation (e.g., availability
of private funds). In other cases, milestone completion involved significant R\&D including reporting experimental data from fusion machines or
prediction of fusion-machine performance via numerical modeling.

Two external ``boards'' provide input and assist the DOE Program Manager,
Dr.~Colleen Nehl. A "Milestone board" provides input on high-level strategy
for the program. An ``input board,'' led by staff members from 
Princeton Plasma Physics Laboratory and Oak Ridge National Laboratory,
assist with milestone review coordination and scheduling, identifying
and securing milestone-specific expert reviewers, and day-to-day operations.
Reviews were/are held either via videoconference or as in-person reviews onsite at company headquarters, depending on the hardware significance of the milestones.

A strong positive of the Milestone Program has been the detailed constructive criticisms and advice provided
by expert reviewers to the awardees, which equates to valuable free advice from a group of the nation's leading experts on a
particular milestone topic. However, this works well only when the advice does not delay DOE decisions
on validating milestone completion, especially if the advice (while still welcomed by the awardees) is clearly beyond the scope of the specific milestone-completion criteria. For example, several awardees stated that there was
scope creep in some of the milestone reviews, where awardees were asked to respond to expansive, open-ended questions 
beyond the specific negotiated milestone-completion criteria.
Another positive has been the value in requiring FPP preconceptual designs early in the program. For example, one awardee
indicated that the preconceptual-design process was highly valuable in identifying requirements and future development needs,
informing applications to the INFUSE program \cite{infuse}, and supporting investor due diligence.

Challenges in program execution include securing enough expert reviewers who are free of conflicts-of-interest (which is also a positive sign that there are strong collaborations among the awardees/companies and the broader fusion public-sector
R\&D ecosystem), lack of timeliness (often tending toward months rather than weeks) in most
administrative matters (including milestone reviews, payments, negotiation of new milestones, and
amendments to award documents), lack of clarity in some cases on the
expected scope or level of detail in milestone reporting (e.g., especially for the FPP preconceptual design) leading to difficulties
for both reviewers and companies, sometimes conflicting direction from DOE to the awardees
on milestone priorities, and recurring uncertainty in the amount of annual budget appropriations to the Milestone Program (which
introduces
challenges/delays for both the DOE and the awardees in defining and negotiating future milestones). Alan Lindenmoyer, former Program Manager for the NASA COTS program who is also on the Milestone board, has repeatedly emphasized the importance of having had \$500 million fully appropriated/available at the outset of the NASA COTS program to facilitate long-term planning and strong signaling to the private sector.

Another significant challenge is the extremely time-consuming process for awardees/companies to partner with DOE or
DOE/NNSA National Laboratories through
Strategic Partnership Projects (SPPs) or Cooperative Research and Development Agreements (CRADAs) \cite{labpartnering}. Key issues 
that we have identified are (1)~it
can and often does take in excess of a year to execute SPPs or CRADAs, (2)~the laboratories have little-to-no flexibilities in negotiating
the SPP or CRADA terms, being that these are dictated by their management and operating (M\&O) contracts with DOE, but
awardees try to negotiate them anyway leading to more wasted time, (3)~foreign-influence reviews are sometimes escalated to DOE headquarters, which adds multiple months to the process with little visibility to
awardees/companies or the laboratories themselves as to progress, and (4)~awardees/companies must repeat the process for each laboratory separately (often for an interrelated scope of work) rather than leveraging much of the review/diligence done already by the first laboratory. All this is further
exacerbated by different laboratories and DOE site offices having different procedures and interpretations of certain rules. While these
challenges are now well-documented by all parties (awardees/companies, laboratories, and DOE) and receiving increasing attention/focus, the challenges are far from solved and raise the
question of whether incremental procedural improvements around the edges will ever be sufficient to effectively support
PPPs such as the Milestone Program and others like it. One thing the IPT did do preemptively, in hopes of saving time
on SPP IP negotiations between awardees/companies and laboratories, was to instruct laboratories to 
adopt the final IP terms of the TIA for any SPPs executed to support the Milestone Program. DOE has additional mechanisms for partnerships between its laboratories
and the private sector that are underutilized, much in the same way that DOE's OT authority has previously been underutilized. 
On example is the Agreements for Commercialization of Technology (ACT) \cite{labpartnering}, which allows certain laboratories to work with private-sector
partners under terms that can be more commercially flexible (including multi-party agreements) than SPPs or CRADAs. This should be further explored and developed,
much in the same way in which we pushed the envelope and the terms under which DOE implements its OT authority.

As of September 2026, the final deliverables for the first 18 months of the Milestone Program are being completed, i.e., awardees' FPP
preconceptual designs and technology roadmaps. Most of the awardees have received notification of completion of these milestones from the DOE, and
four of them have already published special journal issues underpinning their FPP preconceptual designs in the peer-reviewed literature \cite{FPP-T1E,FPP-CFS,FPP-Thea,FPP-TE}. Most of the remaining companies are anticipated to do so in the near future, and 
one has released a standalone paper thus far on its FPP physics basis \cite{frank25}. Of course, protected data/information of the companies will generally not be included
in the peer-reviewed publications. As of September 2026, many companies are presently negotiating their milestones for the second 18-month period of the program even though it is nearly 28 months since TIAs were signed (June 2024).

The question of downselection in the program comes up often. Our view in designing the program to meet its
statutory requirements is that all awardees continuing to meet their milestones, whether toward FPP preliminary
engineering designs or significant performance improvements in their FPP concepts, should have the opportunity
to remain in the program. This supports the development of a U.S. fusion industry and leverages/catalyzes the
greatest amount of private funding into U.S. fusion companies. However, we also believe that most of the program's
budget should increasingly be allocated to the emerging teams that demonstrate (1)~the greatest and most credible
S\&T progress toward realizing FPP preliminary engineering designs by 2030, (2)~the ability to raise the
necessary private capital to construct/operate billion-dollar-scale infrastructure including $Q>1$ machines and
FPPs (expected to be in the U.S., though the FOA did not state this as an explicit requirement), and (3)~commensurate progress toward readiness for commercial demonstration and
deployment (as encapsulated in both Table 5.1 of \cite{NASEM21} and the DOE ARL framework~\cite{ARL}). We
believe there should not be artificial/forced downselections in the program, but that program managers should 
insist on truly meaningful and significant S\&T milestones and make the hard decisions to either cut projects that fail to meet milestones or to allocate commensurately reduced budget
for those making slower progress toward program objectives. We also believe that there should be regular competitive opportunities for new, highly meritorious entrants into the Milestone Program.

\section{Reflections}
\label{sec:reflections}

The DOE {\em Milestone-Based Fusion Development Program} is a first-of-its-kind program for the DOE\@. To our knowledge, it has the
most industry-friendly terms and conditions ever negotiated for a DOE funding agreement, and it was the first time that eight fixed-support TIAs 
were negotiated and awarded in a single DOE program. When juxtaposing the DOE Milestone Program TIA terms against the NASA COTS Space Act Agreement terms (from which we drew inspiration), 
the Milestone Program went further in implementing flexibilities especially with respect to industry-friendly IP terms. The Milestone Program was and remains a flagship initiative for the U.S. policy shift, initiated in 2022, to accelerate fusion commercialization by partnering with the private sector \cite{hsu23,DOE24,FST-roadmap}. 

The Milestone Program stands out as a quintessentially and uniquely American approach amidst a growing landscape of industrial policies pursued by other countries \cite{F4E-report} to promote their own domestic fusion industries. By including a relatively large number of awardees who are able to take on the upfront risk of raising and spending private capital, limiting administrative aspects of program management as much as possible to evaluation/validation of technical milestones, and letting winners who continue to meet all milestones naturally emerge and dropping participants who cannot meet their milestones (while hopefully both avoiding artificial downselections and providing an onramp for new meritorious companies to join the program), the Milestone Program is tailored to the American system. If administered appropriately, it scrupulously avoids picking winners or suppressing underdogs. It does not prop up inefficient or failing enterprises, leaving the ultimate fate of the companies to market and S\&T forces. This also allows federal support of fusion R\&D to be concentrated toward accelerating the most worthy technical and commercialization pathways.

Arguably, the strongest impacts of the Milestone Program thus far have been the affirmation and technical validation of a portfolio of 
privately funded companies as the core of a growing U.S. fusion industry, as well as 
their development plans toward viable FPP designs. The program also serves as a signal of strong government support for accelerating fusion commercialization. Notably, the eight awardees/companies are pursuing five distinct fusion concepts. Even those pursuing the same concept (e.g., stellarator-based Thea and Type One, or tokamak-based CFS and Tokamak Energy) have made highly differentiated design
choices both in terms of physics and engineering. The program elevated the importance of delivering FPP preconceptual designs and technology
roadmaps as early as possible in a company's development path. This focuses attention on the challenges of integrating all
the various subsystems of an FPP, and it helps prioritize resources toward resolving
the most severe S\&T gaps as early and as cheaply as possible.
Expert review of S\&T milestone completion as part of the Milestone Program is intended to be the gold standard for the industry, providing high-quality due diligence for a field that is difficult for investors to evaluate. This not only provides confidence to existing investors but signals credibility to new investors. Interestingly, there has been interest from 
fusion companies to join the Milestone Program as ``zero-payment'' awardees, where
the partnership would be limited only to milestone negotiation, review, and validation with no federal payments nor additional
administrative requirements. We considered offering such ``zero-payment'' awards to the two applicants just outside the eight
that were selected, but we could not get approvals to do so. This was in part due to concerns that it would violate the Congressional authorization of the program and perhaps also raise appropriations legal issues, as DOE staff time would go towards companies that were not selected under the program to receive government funding. A future Congress could consider explicit statutory language allowing for a version of ``zero-payment awards'' for interested companies.

Although the Milestone Program cannot claim/prove direct causality of the
following, we note
that (1) since award selections were announced on May 31, 2023, the ratio of private funding raised by the eight awardees to the program's
obligated federal funding to-date is over 25:1 (not including CFS, the ratio is still over 7:1), and (2) companies have successfully closed significant fundraising rounds
after announcing that DOE validated the completion of significant milestones, e.g., Thea's \$100 million Series B \cite{Thea-fundraise} following
successful completion of their FPP preconceptual-design milestone \cite{Thea-milestone} or the second close (\$1 billion) of CFS' Series B2 \cite{CFS-fundraise} following successful
completion of their SPARC toroidal-field coil test \cite{CFS-magnet}. While the Milestone Program has 
less impact on a company like CFS (which had already raised more than \$2 billion when they joined the Milestone Program) in their private fundraising, earlier-stage companies probably benefit more significantly from being in the program with
respect to private fundraising.

The greatest challenges faced by the Milestone Program have been and continue to be the limited size and uncertainty of the funding, which
limits the effectiveness of the program relative to its policy objectives, as well as the 
entrenched administrative barriers to implementing and executing such a program with the most appropriate terms and rapid decision-turnaround times. We emphasize that the challenges to implementing and negotiating OT-based awards should not be used as a reason
to shy away from their use. One of us, S.H., heard repeatedly from many career staff across the DOE, including ARPA-E, warning
others not to pursue the use of OT agreements due to the difficulties in award negotiations. Based on our experiences, there is nothing wrong with the OT authority/mechanism itself. Instead, challenges and delays were due to award 
negotiations starting from an inappropriate set of terms and conditions and a lack of willingness of AOs, patent counsel, and their chains of
command to exercise the needed flexibilities provided in the statutes when these were clearly needed to fulfill the policy
objectives of the program.

Regarding the size of funding for the Milestone Program, the scale
of mismatch between federal and private funding is only growing.
On the one hand, the high ratio of private-to-federal
funding suggests that the program is successful in amplifying fusion funding and catalyzing large amounts of private funding. On the other hand, the same high ratio is also a result of the modest sums of federal funding that have been appropriated
to the program, which limits the impact of the program and its ability to catalyze private funding as development and capital costs escalate for the awardees as they design, build, and operate billion-dollar-scale facilities.
While federal payments totaling millions to tens of millions of dollars over an 18-month period 
are welcomed by the companies, and in some cases do meaningfully extend a company's capital runway (giving them more time for private fundraising and/or to overcome milestone delays), at some point the administrative burdens of the program may not be justified when overall company expenditures of private funds
are nearly two orders of magnitude higher over an equivalent time period. To continue its role as a catalytic force to amplify fusion funding, the Milestone Program needs federal funding commensurate with a company's level of achieved S\&T de-risking as well as its expenditures over comparable time periods. 
A reasonable point of comparison for the present Milestone Program (i.e., development stage) could be the middle tier of the DOE ARDP, ``Risk Reduction for Future Demonstrations,'' which awarded between
\$85--303 million each to four of the five companies in this tier \cite{ardp} (compare to obligated amounts to-date for Milestone Program that are generally an order of magnitude lower, see Table~2). The greatest risk from the perspective of private fusion investors is
capital formation, and investors are looking for non-dilutive capital/incentives at the level of hundreds of millions of dollars for each of the
leading fusion companies (whose next steps are to construct billion-dollar-scale prototypes) to positively impact and influence a decision to invest in a particular company's funding round. If/when one or more fusion companies passes
their FPP preliminary-engineering-design review, then an appropriate comparison could be the ``Advanced Reactor Demonstrations'' tier of the ARDP, for
which DOE's commitment is reportedly up to approximately \$2 billion to {\em each} of the two companies in this tier (TerraPower \cite{terrapower} and X-energy \cite{x-energy}).
Thus, it would be a mistake to think that the existing amounts of public
funding in the Milestone Program are sufficient to catalyze the additional billion dollars or more that is required for each breakeven-class demonstration or FPP\@.

While the final negotiated TIAs relaxed
many problematic terms and conditions of standard FAR-based contracts, and provided additional mitigations surrounding key awardee concerns 
around reporting burdens, IP provisions, and timelines for administrative actions/approvals, it took too long (1 year) to arrive at an acceptable
TIA template, and much
room for improvement still remains. For example, the starting point of future negotiations for OT agreements could be informed by a 
predetermined matrix
that includes criteria such as stage of RDD\&D (research, development, demonstration, and deployment), policy objectives of the program, fraction of federal funding relative to TPC, and relative degree of upfront financial risk taken by the company versus the government,
etc. We hope that future DOE PPP programs with similar program/policy objectives will leverage our negotiated TIA
as a starting point rather than an ending point. In addition, for implementing/executing
PPP programs such as the Milestone Program, it would be worth introducing appropriate incentives for federal staff to
deliver decisions on a pace and with risk tolerance that is compatible with the requirements of venture-backed, frontier-technology companies, and to consider standing up a separate implementation/approval process that is entirely independent of the existing DOE Financial Assistance regime (which has been optimized for grants and cooperative agreements with predominantly or 100\% federal funding).

Finally, we highlight that this program could be leveraged as a ``template'' for other milestone-based PPP programs. For example, our efforts in designing
the Milestone Program have already benefited the ARDP in their own efforts to award fixed-support, milestone-based OT agreements with select ARDP awardees.
With minor modifications
to the Milestone Program TIA, it could be utilized for a variety of PPP programs, spanning the full spectrum of RDD\&D, that seek to partner with the private sector. In particular
for fusion, the same TIA template (hopefully with continued improvements based on experiences with the present Milestone Program and 
feedback from the awardees) could be used for industry/consortia-led construction and operation of fusion materials and technology (FM\&T) test stands (as
described in the DOE Fusion Science and Technology Roadmap \cite{FST-roadmap}), as well as a future demonstration tier of the Milestone Program, i.e., a {\em Milestone-Based Fusion Demonstration Program}.

\backmatter

\bmhead{Acknowledgements}
We acknowledge many people, presently or formerly with the DOE, who contributed directly to the
design and/or implementation of the DOE Milestone Program. For design,
we acknowledge Rich Hawryluk.
For implementation, we acknowledge Colleen Nehl, John Mandrekas, Yasmin Yacoby,
Cynthia Adams, Warren Riley, Jennifer Harling, Michael Dobbs, Michael Goldstone (deceased), Brian Lally, Jennifer Mahalingappa,
Michael Zarkin, and Derek Passarelli. We thank
Rich Hawryluk, Colleen Nehl, and several of the Milestone awardee points-of-contact for providing detailed feedback on the manuscript. Finally, we thank all the awardees for your strong partnership in the Milestone Program.

\section*{Declarations}

S.H., Fusion Partner at Lowercarbon Capital, formerly DOE Lead Fusion Coordinator (2022--2025) and ARPA-E Program Director (2018--2022), led the design
and supported the award negotiations, implementation, and early execution of the DOE Milestone Program.
S.W., founder of Fusion Energy Base, formerly an ARPA-E Technology-to-Market Advisor (2021--2024), supported the design, award negotiations, and implementation of the DOE Milestone Program. N.S., Non-Resident Fellow at the Center on Global Energy Policy at Columbia University, formerly Advisor to the Secretary of Energy (2023--2024)
and Legal Advisor in the DOE Office of the General Counsel (2021--2023), supported the award negotiations and implementation of the DOE Milestone Program.

{\bf Competing interests.} The authors had no financial interests in any of the companies named in this paper while they were employed by the DOE, when this work was performed. 
S.H. and S.W. now have financial interests in some of the companies named in this paper, and some of the companies named in this paper are clients of the employer of N.S.

\begin{appendices}
\setcounter{table}{2}

\section*{Appendix: Summary Table and Distilled Lessons Learned}\label{appendix}

This appendix provides a summary (Table~\ref{tab:policy_design_challenges}) of program objectives, design features, and
implementation challenges, as well as a distillation of lessons learned
(organized by section), as drawn from the entire paper.

\begin{sidewaystable}
\caption{Summary of the Milestone Program's objectives, program design features, and implementation challenges.}
\label{tab:policy_design_challenges}
\footnotesize
\renewcommand{\arraystretch}{1.3}
\begin{tabularx}{\textheight}{>{\hsize=0.5\hsize\raggedright\arraybackslash}X >{\hsize=1.25\hsize\raggedright\arraybackslash}X >{\hsize=1.25\hsize\raggedright\arraybackslash}X}
\toprule
\textbf{Program objective} & \textbf{Program design feature} & \textbf{Implementation challenge} \\
\midrule
Support the development of a competitive U.S. fusion industry and accelerate fusion commercialization via PPPs that leverage and further catalyze private investments.
&
Used OT authority to maximize compatibility with venture-backed startups willing to take significant upfront financial risk, where startup-friendly IP terms and reduced reporting requirements could be negotiated. Cost-share waiver eliminated CAS and enabled use of fixed-support TIAs, which avoids burdensome line-item budget reporting ($>50$\% non-federal contribution still required). Final TIA terms included much narrower government data and IP rights, with extended protections compared to FAR-based contracts. Merit-review process evaluated S\&T, business/financial, and commercialization criteria with equal weighting, leading to selection of all eight highly meritorious applicants (representing five distinct fusion concepts) to maximize probability of program success and amount of private capital catalyzed. In general, ``insight not oversight'' was the aspiration.
&
DOE award negotiations and approval processes were built for awarding FAR-based contracts rather than OT agreements with venture-backed startups needing flexibility/speed. Lack of experience in awarding TIAs to venture-backed startups by DOE AOs/patent counsel and competing policy directives (especially relating to FAR-based IP treatment and degree of oversight
on U.S. manufacturing and research security) led to inappropriate starting point for negotiations, requiring Secretary-level intervention to move past negotiations stalemate after six months. Federal contributions being one-to-two orders of magnitude smaller than awardees' private expenditures over a similar time period limits the program's impact in accelerating fusion commercialization and catalyzing greater private investments, especially as companies progress toward construction and operation of billion-dollar-scale R\&D infrastructure.\\
\midrule
Support awardees' S\&T de-risking toward FPP preliminary engineering designs, including facilitating partnerships with publicly funded research institutions on broad technical scope.
&
Ability to negotiate milestones that are aligned with awardees' investor milestones, emphasis on critical-path S\&T milestones with quantitative targets, required early delivery of FPP preconceptual designs and technology roadmaps, and support of SPP/CRADA negotiations between awardees and national laboratories with directive to the latter to use the TIA's more
industry-friendly IP terms.
&
Final negotiated data rights and IP terms, despite being significantly more industry-friendly compared to FAR-based terms, resulted in many awardees limiting their milestone technical scope and deliverables to avoid government IP and data rights. SPP/CRADA processes caused significant delays (typically $>6$ months and sometimes $>12$ months) and had to be repeated for each national laboratory, leading to delayed project starts and milestone completion in some cases.\\
\midrule
Support awardees' business and broad commercialization objectives as fusion developers.
&
In addition to above, aimed to further streamline waiver processes with respect to U.S. manufacturing and foreign work with promised decisions in 45 days, and limited "flow-down" reporting requirements (especially with respect to foreign influence) to subawardees.  Offered data rights protection for up to 30 years and safe harbor from exercise of government IP rights for an extended period of time after milestone completion. Required milestones on community benefits/engagement (to support siting and workforce development) and overall adoption readiness.
&
Award negotiations and final TIA terms and conditions for this program left open several unresolved questions with respect to the most optimum set of terms for a PPP to support venture-backed startups building a new U.S. fusion industry who are
engaged in a global race with global consequences. In particular, requirements on U.S. manufacturing, research security, and foreign influence may warrant a more graded approach rather than a one-size-fits-all implementation.\\
\bottomrule
\end{tabularx}
\end{sidewaystable}

\subsection*{Origins}
\begin{itemize}
\item It took many years of parallel groundwork (industry advocacy, ARPA-E attention on fusion commercialization, authorizing
legislation, consensus expert reports, and first budget appropriations) culminating in a White House Summit to
finally catalyze DOE to launch the program in 2022.
\end{itemize}

\subsection*{Objectives and Design}
\begin{itemize}
\item The program's multi-faceted objectives led to downstream tensions, e.g., leveraging fast-paced, venture-backed
fusion companies to accelerate fusion commercialization and wanting them to benefit from R\&D partnerships with DOE
national laboratories with extremely slow processes for executing SPPs/CRADAs.
\item OT authority was legislated and fixed-support TIAs were chosen for their theoretical flexibilities, but the
willingness/ability of AOs, patent counsel, and their chains of command to actually exercise them went untested until award negotiations began
\item The appropriated budget
remains an order of magnitude short of program needs, limiting the program's ability to move rapidly and with certainty.
\end{itemize}

\subsection*{Application Process}

\begin{itemize}
\item The fast timeline for FOA development and release (under five months including a workshop) required senior DOE
leadership attention and prioritization
\item Deviating from FAR-based terms required numerous bespoke, high-level administrative approvals before the FOA could even be released (and also later during award negotiations)
\item Application demand/quality outstripped the budget with eight highly meritorious applicants against 
anticipated three-to-five awards.
\item Foreign-influence and research-security reviews were not done in parallel with merit review of applications
and delayed selections by over a month.
\end{itemize}

\subsection*{Award Negotiations}

\begin{itemize}
\item Despite alignment on policy objectives within the IPT, AOs and patent counsel were not experienced in
implementing OT agreements and began from standard FAR-based terms, assuming faster internal approvals; the six-month stalemate that followed, resolved only via Secretary-level intervention, demonstrated the difficulties of
deviating from default practices
\item Defining milestones with quantitative performance targets that represent significant S\&T de-risking toward an FPP required iteration with DOE program staff during award negotiations
\item Releasing the FOA before a draft TIA template was developed with industry input (to preserve momentum) contributed to the one-year negotiation that followed; a small federal share further gave selectees little reason to accept FAR-like terms as a starting point for negotiations
\item DOE's standard reporting checklist, built for federal-majority funding to institutions used to
contracting with the government, does not easily scale down to a program where venture-backed companies supply most of the funding; mutually acceptable terms emerged only through extensive, ad-hoc, line-by-line compromise
\item Starting from DOE's standard IP provisions, designed for federal-majority funding, was a poor fit and a major delay driver; even after negotiated flexibilities, companies still de-scoped S\&T milestones to avoid remaining IP entanglements
\item Without a defined turnaround commitment, waiver and research-security-review timelines were unpredictable and open-ended by default, a poor fit for venture-paced companies; a fixed 45-day clause emerged only as a late compromise
\item The same reactive 45-day compromise resolved the immediate timeline problem for approvals of U.S.-manufacturing waivers, but left unresolved the deeper tension between R\&D-phase domestic-manufacturing mandates and companies' eventual need for globally competitive manufacturing and fusion-plant deployments.
\end{itemize}

\subsection*{Initial Program Execution}
\begin{itemize}
\item Conflict-free expert reviewers were difficult to secure, and administrative turnaround (milestone reviews, payments, TIA amendments) tended toward months rather than weeks, compounded by recurring year-to-year appropriations uncertainty (unlike NASA COTS's fully appropriated funding at outset)
\item Though most awardees stated that they received extremely valuable constructive criticisms during milestone
reviews from the expert reviewers, in many cases, they felt that the reviewers exceeded their mandate by asking
expansive, open-ended questions beyond the scope of the milestone-completion criteria that led to delays in DOE's
validation of milestone completion
\item SPPs/CRADAs with national laboratories became a major, separate bottleneck sometimes exceeding a year, constrained by inflexible DOE/Lab M\&O contracts, escalated foreign-influence reviews, and no reuse of one lab's review by another, while a 
theoretically more flexible and appropriate existing mechanism (ACT) was generally not considered
\item The most visible deliverables from the first 18 months of the program, FPP preconceptual designs
and technology roadmaps, have emerged with many awardees publishing extensive collections of papers on their FPP preconceptual designs in peer-reviewed journals.
\end{itemize}

\subsection*{Reflections}
\begin{itemize}
\item The Milestone Program, to our knowledge, has the most industry-friendly terms and conditions ever negotiated
for a DOE funding agreement; nevertheless, especially from the perspective of the awardees, there is still misalignment between
the policy objectives of the program (primarily to build a competitive U.S. fusion industry and to accelerate fusion commercialization via PPPs) and the final negotiated TIA terms
\item That reaching a mutually acceptable TIA took a year, despite the necessary statutory flexibility (10 CFR 603) already existing, shows regulatory flexibility is necessary but not sufficient; agency culture, incentive structure, and risk tolerance largely
governs willingness and ease of implementation; the present DOE contracting structure and approval processes were/are not
well suited to the implementation of the Milestone Program nor its continued execution
\item The program's $>25$:1 private-to-federal-funding leverage ratio is double-edged: it signals catalytic success but also that federal contributions are too small to meaningfully move company decisions as costs escalate; no predetermined framework existed for calibrating award terms to the program's policy objectives, level of private funding, or the RDD\&D stage
of the work
\item Despite the challenges of negotiating the final terms of the TIA for the Milestone Program, it could be used
(hopefully with improvements based on the lessons learned from this paper)
as the starting point for other similar PPP programs, including R\&D infrastructure and demonstration-phase projects
(i.e., FPPs) in fusion.
\end{itemize}




\end{appendices}



\end{document}